\documentstyle[11pt,newpasp,twoside,epsf]{article}
\def\edcomment#1{\iffalse\marginpar{\raggedright\sl#1\/}\else\relax\fi}
\reversemarginpar

\begin{document}
\title{CDM in  LSB Galaxies: Toward the Optimal Halo Profile}
 \author{W.J.G de Blok}
\affil{Department of Physics and Astronomy, Cardiff University, 5 The Parade, Cardiff, CF24 3YB, UK}

\begin{abstract}
Low Surface Brightness (LSB) galaxies are dominated by dark
matter. High-resolution rotation curves suggest that their
total mass-density distributions are dominated by constant density
cores rather than the steep and cuspy distributions found in Cold Dark
Matter (CDM) simulations.  The data are best described by a model with
a soft core with an inner power-law mass-density slope $\alpha =
0.2 \pm 0.2$. However no single universal halo profile provides an
adequate description of the data. The observed mass profiles appear to
be inconsistent with $\Lambda$CDM.
\end{abstract}

\section{Introduction}
Cosmological numerical Cold Dark Matter (CDM) simulations predict a
specific and universal shape for the dark matter density distributions
of galaxies. This was investigated in detail by Navarro, Frenk \&
White (1996, 1997) [NFW] who found that the mass-density distribution
in the inner parts of simulated dark-matter-only CDM halos could best
be described by a $r^{-1}$ power law. This is however not what is
typically observed: the mass-density distributions of, for example,
dwarfs and late-type galaxies (which do include baryons) are best
described by an almost constant density central core $\rho(r) \sim
r^{-0.2 \pm 0.2}$ (de Blok et al.\ 2001b).  This disagreement has become
known as the ``cusp/core problem''.

There has been much discussion of the validity of both observations
and simulations. My intention here is not to give an exhaustive review
of this debate. Rather I want to highlight some of the concerns and
criticisms of both observers and simulators, and discuss them from an
observer's point of view (and in the spirit of the Sydney panel
discussion I will do this in a slightly provocative manner).

\section{The Data}

\subsection{Consistency and Repeatability}

The rotation curves of LSB galaxies are some of the primary evidence
for the failure of the NFW model in the centres of galaxies.  How
reliable are these data? An often heard complaint is that rotation
curve observers cannot seem to agree amongst themselves on their own
data. Unfortunately, this is occasionally used as an excuse to dismiss
these data. This apparent disagreement is mostly a sociological effect.
A large part of it can be traced back to spin (where I use the
non-astronomical definition of the word) and over-reliance on a few
discrepant rotation curves that have usually already been rejected by
the observers.

\begin{figure}[t]
\plottwo{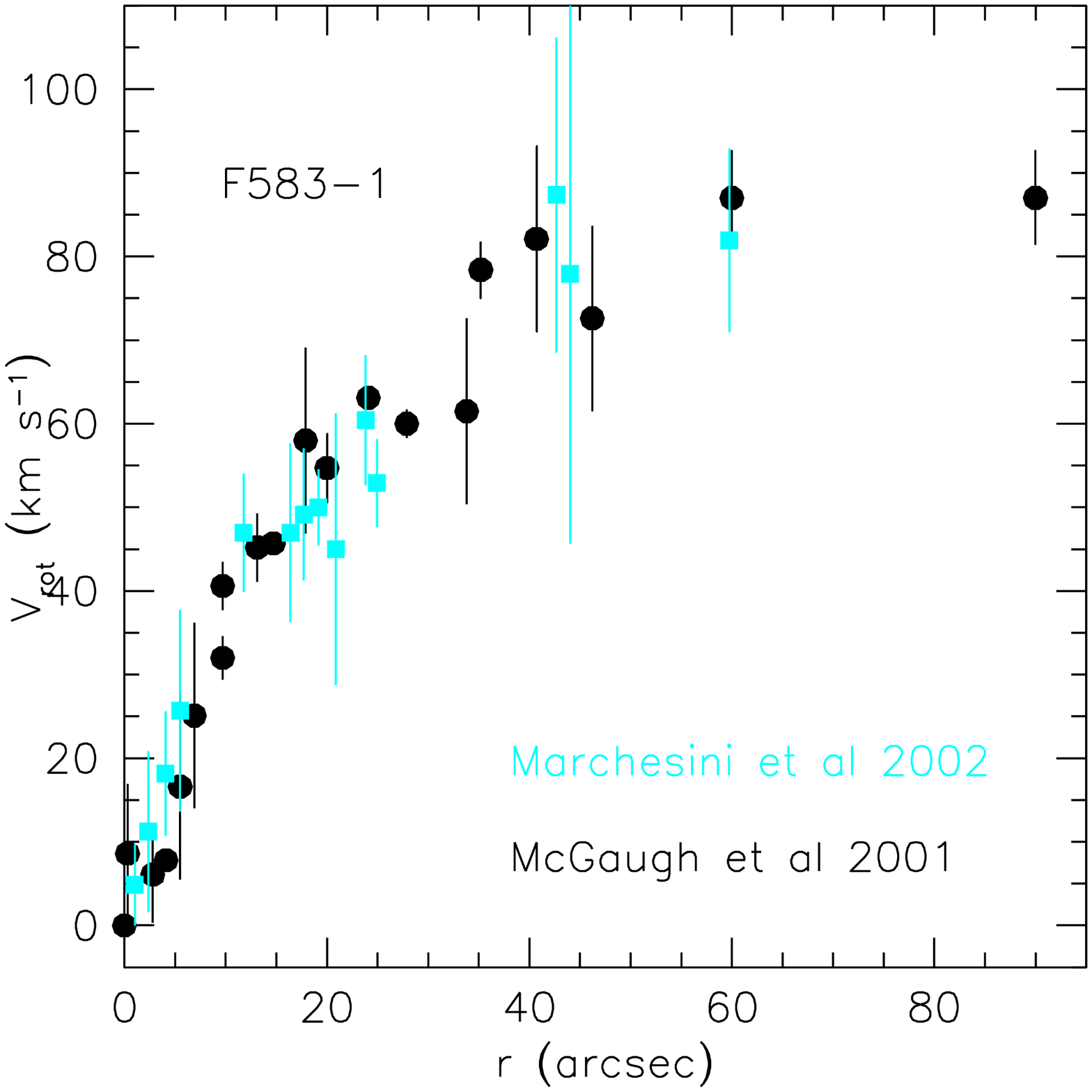}{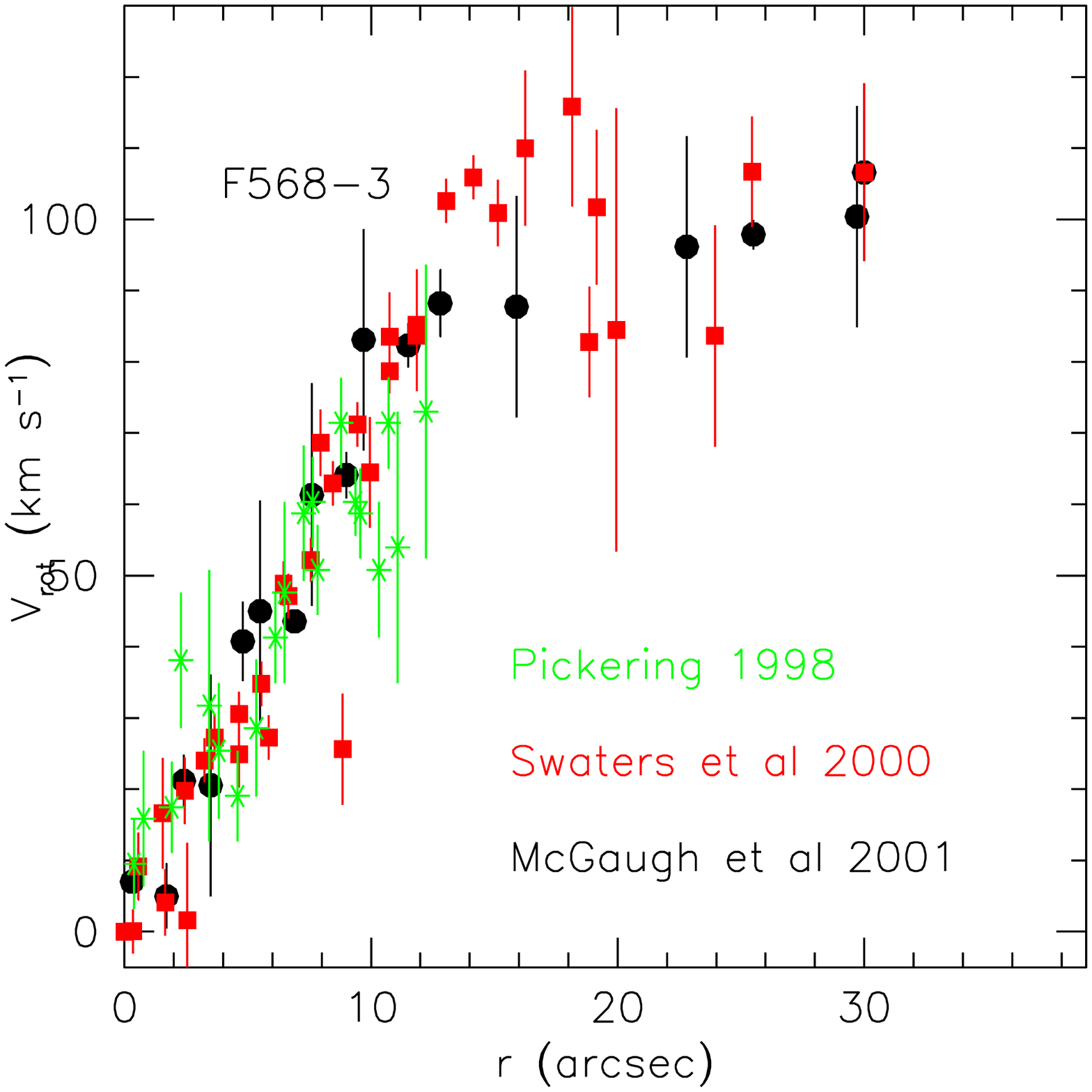}
\caption{Comparison of raw H$\alpha$ rotation curves observed by independent
groups. \emph{Left panel:} F583-1. Circles: data from McGaugh, Rubin \& de Blok (2001). Squares: data from Marchesini et al.\ (2003).
\emph{Right panel:} Circles: data from McGaugh et al.\ (2001). Squares: data from Swaters, Madore \& Trewhella (2000). Asterisks: data from Pickering (1998).}
\end{figure}

Systematic effects due to pointing of the telescope or beam-smearing
also feature prominently.  In the latter case the few curves where
beam-smearing was at the time a genuine problem have received a great
deal of attention (e.g.\ Swaters et al.\ 2000), in contrast with the
dozen or so where the problem was insignificant (de Blok et al.\ 2001a,
de Blok \& Bosma 2002).  Beam-smearing was and is a red herring.

Similar conclusions apply to the pointing ``problem''. Here the
observed shallow slopes are thought to be caused by the spectrograph
slit not being precisely centered on the galaxy. This would then cause
an NFW halo to mimic a halo with a shallow profile.  The problem is
that one would have to miss the centre of a galaxy with an NFW halo by
an average of $\sim 4''$ or so to observe an apparently shallow
profile (de Blok, Bosma \& McGaugh 2003). In reality the centre of a
galaxy can be acquired with an accuracy of $\sim 0.3''$, an order of
magnitude better.  In addition, if NFW halos are really prominently
present in LSB galaxies one would have to miss the centres of these
galaxies by a factor of 10 more than the pointing error for
\emph{each} and \emph{every} of the over 80 galaxies observed. If
pointing errors are not the problem, but rather offsets between the
dynamical and photometric centres of galaxies, then the $4''$ value
translates into a physical offset of $\sim 1.4$ kpc (for the galaxies
in de Blok et al.\ 2003). This would mean that in many LSB and dwarf
galaxies optical and dynamical centres would be separated by roughly a disk
scalelength!

In other words, pointing effects, though of course present in the data
to a \emph{small} degree, are by no means enough to challenge the
observations.  As a picture can say more than a thousand words, we
illustrate this by comparing rotation curves of LSB galaxies observed
by different independent groups at different times with different
telescopes.  There is now enough data in the literature that this can
be done for over a dozen LSB galaxies. Space does not allow us to
present all galaxies here (see de Blok et al.\ 2003 for more
comparisons), but we illustrate with two examples: LSB galaxies F583-1
and F568-3 (Fig.~1). The latter has even been observed by three
independent groups. To avoid confusion due to mixing of H$\alpha$ and
HI data (as discussed below) we compare the raw H$\alpha$ data only.
It is obvious that within the error bars the data agree beautifully.
Many of these curves are now starting to be confirmed by
high-resolution 2-D velocity fields (e.g., Chemin et al., this
volume).  This would be quite a fluke if systematic errors really were
dominant.  In summary, \emph{the data are consistent.}

\subsection{Likelihood}

If the overwhelming majority of observational data are consistent with
each other and therefore with dark matter cores, how does this explain
the often-cited statement usually attributed to Swaters et al.\ (2003)
that LSB rotation curves are consistent with NFW halos?  The Swaters
et al.\ analysis actually does not claim this in such strong terms.
Its strongest statement in this respect is that 75\% of the galaxies
in their sample are ``consistent'' with a NFW model. What this means
is made clear at the end of the Swaters et al.\ paper:
\emph{``Even though the majority of the galaxies in this sample seem
consistent with steep inner slopes, none of the galaxies
\underline{require} halos with $\alpha = -1$, as most galaxies are
equally well or better fitted by halos with shallower density profiles
or even constant density cores.''} How one interprets this statement
depends only on whether one prefers the mode of a distribution as
representative (without the need to invoke systematic effects), or the
extremes of that distribution along with an epicyclic range of
systematics.  Here Occam's Razor has often proved to be invaluable.

\subsection{Clean Sample}

The rotation curves in the literature span a large range in
resolution, inclination, and general quality. It is therefore worth
asking whether the data may in any way be affected by the effects of
bars, non-minimum disc situations, resolution effects, etc.  We thus
isolate the highest quality rotation curves from the total sample from
de Blok et al.\ (2001a) and de Blok \& Bosma (2002) (the open
histogram in Fig.~2 [left panel]). We have applied the following cuts: (1) we
remove all galaxies with inclinations $i < 30^{\circ}$ and $ i
>85^{\circ}$ (single hatched histogram); (2) We remove all galaxies
with low-quality rotation curves, i.e.\ with a small number of
independent data points, with large error-bars, and large asymmetries
(double hatched histogram); (3) Galaxies which were clearly not
dominated by dark matter (e.g.\ with bulges) were removed; (4) As the
difference between cusp and core is most clearly visible in the
innermost part of the galaxies, we demand that the inner part is
resolved: we retain only those galaxies that have at least 2
independent data points in the inner 1 kpc (grey filled
histogram). This leaves us with a final sample of 19 galaxies, out of
over 80 originally.

\begin{figure}[ht]
\plottwo{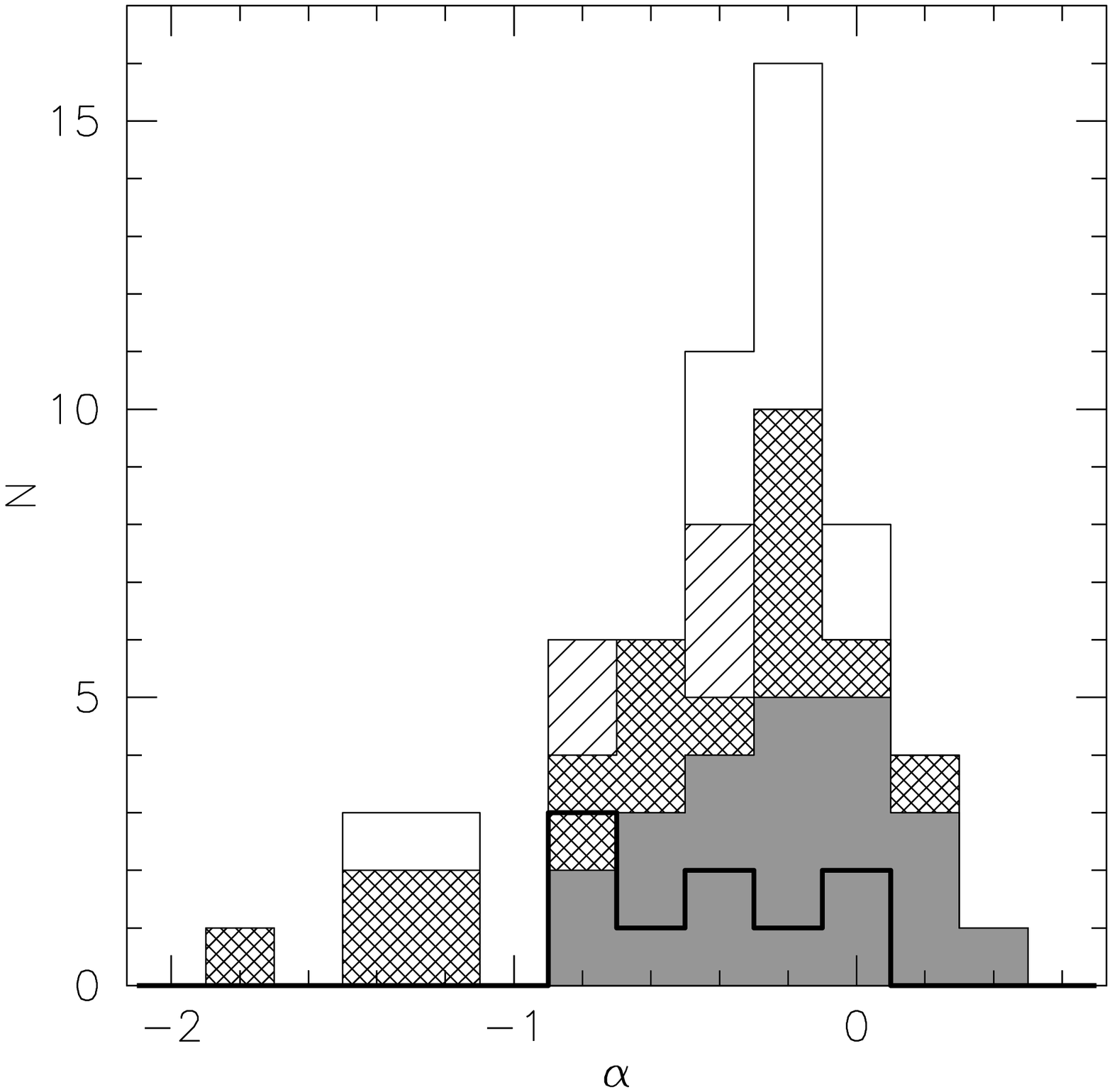}{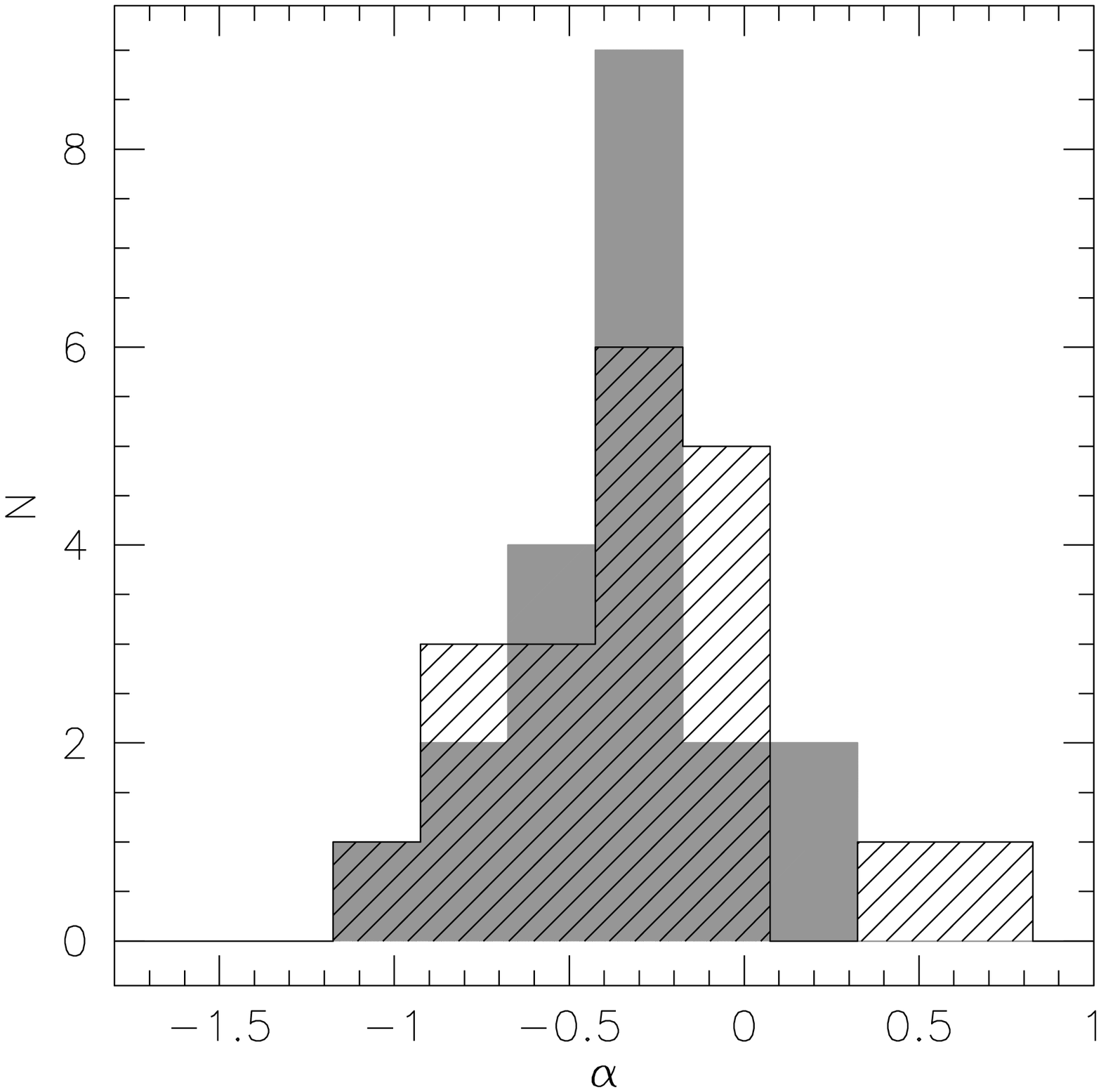}
\caption{\emph{Left:} Histogram of inner mass-density slopes. 
See text for explanation. \emph{Right:} Distribution of the
mass-density slopes of galaxies from de Blok et al.\ (2001a) for which
photometry and HI are available.  Full grey histogram: minimum
disc. Open, hatched histogram: non-minimum disc, $M/L_*(R)=1.4$.}
\end{figure}

A comparison of the different stages of pruning shows no systematic
differences between the distributions of the initial full sample and
final restricted sample. If anything, the peak of the distribution of
inner slopes $\alpha$ has shifted towards a more positive value. The
tail towards steep negative slopes has disappeared.

As described above, an analysis by Swaters et al.\ (2003) finds that
the slopes in their sample cover the full range $-1
\la \alpha \la 0$, with a preference for  shallow
slopes. Below we show that a large fraction of the steep slopes are
actually due to insufficient resolution.  We have thus subjected their
data to the same quality criteria that we used for ours. The results
are over-plotted in Fig.~2 (left panel; thick line).  The number of remaining
galaxies in the Swaters et al.\ sample is small, but it is clear that
the distribution is entirely consistent with our larger restricted
sample.

Our restricted sample has three galaxies in common with their
restricted sample. We find that for two of these galaxies the measured
slopes are consistent at better than $1\sigma$ (UGC 731 and UGC
4325). For one galaxy (UGC 11557) the slopes differ significantly:
they find $\alpha=-0.84 \pm 0.27$, while we find $\alpha=-0.08 \pm
0.23$. An inspection of the density profiles in Fig.~1 of de Blok et
al.\ (2001b) and Fig.~5 in Swaters et al.\ (2003) shows that this
difference in slope is entirely due to the range over which the slope
is measured. The data themselves are consistent.  Swaters et al.\
(2003) have chosen a larger fitting range, resulting in a much steeper
slope. Choosing a smaller range results in identical slopes.  This
comparison of restricted samples shows that once a proper pruning is
made to only include minimum-disc, well-resolved galaxies, most of the
differences are merely a matter of semantics. The observed
\emph{shallow} inner mass-density slopes of LSB and dwarf galaxies do
therefore reflect these galaxies' potentials.

\subsection{Minimum disc}

As described in de Blok et al.~(2001b) the inversion from rotation
curve to mass-profile is only valid in the limit of spherical halos
consisting only of dark matter. The last assumption is known as
``minimum disc'' and has been used in all rotation curve inversions
described here.  The minimum disc assumption ascribes all dynamics and
all mass to the dark matter component, and is thus really a ``maximum
halo'' model. As such it gives strong upper limits on the measured
dark matter slopes.

A proper treatment of non-minimum disc models should include the
influence that dark and visible matter have on each other in the disc,
but to first order we can derive the rotation curve under a
non-minimum disc assumption and invert this curve. This should give an
indication of the change in inferred dark matter slopes when going
from a one-component to a two or three component mass model.

We have done this for the rotation curves in de Blok et al.\ (2001a)
for which photometry and HI data were available, and have derived the
halo rotation curves assuming $(M/L)_*(R) = 1.4$.  The dark matter
mass-density slopes found in this way are shown in Fig.~2 (right
panel), where they are compared with the minimum-disc slopes for the
same galaxies. The peak of the slope histogram is still found near
$\alpha=-0.2$, showing that LSB galaxies are indeed dark matter
dominated. The bulk of the distribution has shifted to more positive
slopes, making the discrepancy with the predicted $\alpha=-1$ value
larger. It is important to realise that in more sophisticated models
one has to take into account that the halo reacts to the presence of a
disc by contracting slightly. This means that the original
``cosmological'' dark matter slopes must have been \emph{even less
steep} than the ones currently measured.

\section{Comparing with simulations}

\subsection{Apples and Oranges}

Despite the overwhelming evidence that measuring rotation curves is a
reliable and repeatable exercise, claims do sometimes appear in the
literature suggesting the opposite.  In every one of these cases
though, not all aspects of the data and their limitations are taken
into account, which often leads to comparing apples with oranges.


A comparison of two datasets of UGC 4325 in Hayahshi et al.\ (2003)
suffers from this.  One of them, from de Blok
\& Bosma (2002), is a \emph{processed (smoothed) H$\alpha$-only} rotation 
curve.  The second data set is taken from Swaters et al.\ (2003) and
consists of \emph{raw H$\alpha$ data combined with processed HI
data}. Hayashi et al.\ (2003) compare both sets and point out the
different nature of the error bars in both sets (as one would expect
given the different natures of the data sets; however, it leads them
to reject the use of $\chi^2$ in data analysis).  They fail to take
into account that the outer points of the Swaters et al.\ rotation
curve are determined by HI, not H$\alpha$, which leads to an entirely
different weighting of the data.  All this leads to an inappropriate
and apparently unfavourable comparison.

Comparing like with like, for example by comparing the raw Swaters et
al.\ H$\alpha$ data with the raw de Blok \& Bosma H$\alpha$ data we
see none of these glaring disagreements (Bosma, this
volume). Similarly, a comparison of the raw de Blok \& Bosma H$\alpha$
data with the raw Swaters et al\ HI data (as shown in de Blok
\& Bosma 2002) also shows good agreement. 

The lesson to be drawn from this, is that data are not as
black-and-white as theoretically motivated comparisons sometimes
portray. Data are not just ``good'' or ``bad''; they may be both,
depending on the situation in which they are used.  It is vital to be
aware of the limitations.  In every large collection of data it is
always possible to find the odd one out (which is precisely why one
collects large data sets). For LSB rotation curves, when proper
comparisons are made, agreement is almost always found.  This may
explain why virtually all observational papers agree on the
inappropriateness of the NFW fitting formula for describing LSB galaxy
dynamics (contributions by, amongst others, Bolatto, Chemin, Zachrisson, Pizella, Gentile, this
volume).

\subsection{The limit of simulations}

In this context, it is instructive to compare the new high-resolution
simulations by White and Navarro (this volume) and Hayashi et al.\
(2003) with observations.  These authors present new high-resolution
simulations of dark matter halos in a $\Lambda$CDM universe. These are
now becoming just resolved enough to start probing the regime of dwarf
and LSB galaxies, however, they also show the large gap that still
exists between simulations and observations.

Observationally, we can plot the inner mass density slope $\alpha$
against resolution $r_{in}$ of a rotation curve. Such a diagram
(Fig.~3) clearly shows that at high resolutions ($r_{in} \la 1$ kpc)
the core and cusp model can be clearly distinguished from each other,
but that that at low resolutions $r_{in} \sim 1$ kpc the cusp model
and the core model have identical slopes (de Blok et al.\ 2001b).
Using Fig.~4 in Hayashi et al.\ (2003) we have computed the values
$r_{in}$ for their simulations and show the corresponding slopes in
Fig.~3.

\begin{figure}[t]
\plotfiddle{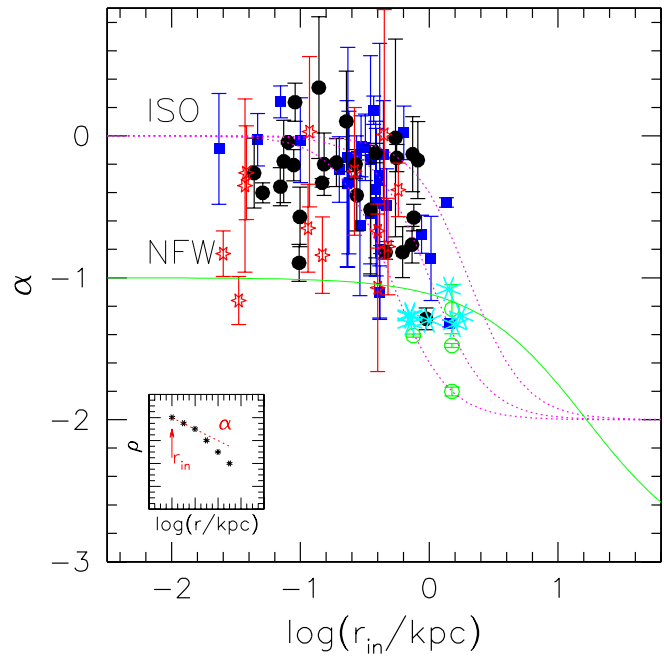}{8.5cm}{0}{45}{45}{-130}{-65}
\caption{Inner mass-density slope $\alpha$ versus resolution $r_{in}$ of
the LSB rotation curves. Symbols with error bars are observational
data.  Circles: de Blok et al.\ (2001a); squares: de Blok \& Bosma
(2002); open stars: Swaters et al.\ (2003). The large asterisks near
$\alpha \sim -1$, $r_{in} \sim 0$ are the simulations by Hayashi et
al.\ (2003). Curves indicate predicted slopes for various core models
(dots) and a NFW model (full line).  See de Blok et al.\ (2001b) for a
full description.}
\end{figure}

The smallest scales these new simulations resolve are $\sim 1$ kpc,
which is exactly the scale below which the core/cusp discrepancies
become apparent. For now, the current models fall solidly in the
range where the core and cusp models make identical predictions.

Going back to the original NFW papers (NFW 1996,1997) one sees that
these older simulations also did not resolve scales smaller than 1
kpc. However, the discussion in these papers sometimes make it all too
easy to forget this limitation.  For example, NFW (1996)
states `The density profiles of CDM halos \emph{of all masses} can be well
fitted by an appropriate scaling of a ``universal'' profile \emph{with no
free shape parameters}' (emphasis added) and the same paper presents an analysis of NFW
rotation curves comparing them directly to those of dwarf galaxies.
One can see how they would have
provided an impetus for the cusp/core debate.

Simulations thus need to improve by another order of magnitude in
resolution before the predictions become relevant for the
observational regime currently probed. In summary, current simulations
do not yet tell us anything about the nature of dark matter at scales
$\la 1$ kpc.

\section{The future}

The study of the distribution of dark matter within the visible discs
of galaxies is now clearly driven by observations. Increases in
computing power will at some stage resolve the inner kpc in
simulations as well, but given the ``coldness'' of CDM and its lack of
cross-section it is unlikely that the slopes will decrease in the
\emph{dark-matter-only} simulations (unless of course the properties
of the dark matter particle are changed).  It is thus interesting to
note that the problems of CDM only become clear on length scales where
the baryons start playing a role (this applies to both the cusp/core
problem as well as the missing dwarfs problem). This indicates that
baryon physics is one of the missing pieces of the puzzle, and will
very likely make a major contribution toward a solution (see, for
example, Sancisi, this volume).

Can observations of galaxy dynamics at scales larger than a kpc, where
the simulations do have predictive power and baryons may be
unimportant, help resolve the issue? The answer to this is yes and
no. CDM makes a robust prediction that the density of matter should
drop off as $r^{-3}$ at large radii in galaxies.  However, finding
such a drop-off is not so much a triumph for CDM, as it is a triumph
for gravity. The $r^{-3}$ behavior of dark matter at large scales is
simply the behaviour of \emph{any} pressure-less medium under the
influence of gravity. The success of large-scale structure simulations
has shown that at large scales this is exactly how dark matter
behaves.  An $r^{-3}$ drop-off does \emph{not} tell us how dark matter
behaves at small scales and large densities. Maybe the observed dark
matter cores are telling us that dark matter has pressure at small
scales. Maybe dark matter is not cold at these scales. Maybe they tell
us something unexpected.  The centers of galaxies are the only places
where we can study dark matter under these conditions. It would thus
be unwise to ignore the conclusions forced upon us by the data.

\section{Conclusions}

\begin{itemize}

\item The available data overwhelmingly prefer soft cores.

\item There is a large range in inner slopes. Observationally there is
no evidence for a ``universal profile''.

\item Systematic effects play no significant r\^ole.

\item Observations probe a regime where simulations have no predictive power (yet).

\item Crucial (baryon) physics are still missing from the simulations.

\end{itemize}


\end{document}